\documentstyle[11pt]{article}
\begin{document}
\vspace{-20ex}
\vspace{1cm}
\begin{flushright}
\vspace{-3.0ex}
    {\sf ADP-99-11/T356} \\
\vspace{-2.0mm}
\vspace{5.0ex}
\end{flushright} 

\centerline{\bf\large NON-SPECTATOR DIQUARK EFFECTS ON LIFETIMES OF
$\Lambda_b$, $\Omega_{b}^{(*)}$}
\vspace{0.5cm}
\centerline{\bf\large  AND WEAK DECAY RATES OF $\Sigma_{b}^{(*)}$,
$\Xi_{b}^{(*)}$}

\vspace{1cm}

\centerline{Wu-Sheng Dai$^{1,2}$, Xin-Heng Guo$^{3,4}$, Xue-Qian Li$^{1,2}$
and Gang Zhao$^{1,2}$}
\vspace{1cm}
\centerline{1. CCAST (World Laboratory), P.O. Box 8730, Beijing 10080,
P.R. China}
\vspace{0.5cm}
\centerline{2. Department of Physics, Nankai University, Tianjin 300071,
P.R. China}
\vspace{0.5cm}
\centerline{3. Department of Physics and Mathematical Physics,}
\centerline{and Special Research Center for the Subatomic Structure of
Matter,}
\centerline{University of Adelaide, SA 5005, Australia}
\vspace{0.5cm}
\centerline{4. Institute of High Energy Physics, Academia Sinica, Beijing
100039, P.R. China}
\vspace{0.5cm}

\vspace{0.5cm}

\vspace{1cm}

\begin{center}
\begin{minipage}{12cm}
\noindent{\bf Abstract}

The difference of $\tau_{_B}$ and $\tau_{_{\Lambda_b}}$ indicates the role of
the light flavors. We calculate the lifetimes of B-meson and $\Lambda_b$
based on the weak effective Hamiltonian while assuming the heavy baryon is
constructed by a heavy b-quark and a diquark containing two light quarks. In
this scenario, we use the information of the measured ratio
$\tau_{_{\Lambda_b}}/\tau_{_B}$ as input to predict rates of the inclusive
weak decays of $\Sigma_{b}^{(*)}$ and $\Xi_{b}^{(*)}$
into non-bottom final states.
We find that these rates of $\Sigma_{b}^{(*)}$ and $\Xi_{b}^{(*)}$ are much
larger than that of B-mesons and $\Lambda_b$. We also give the predictions
for the lifetimes of $\Omega_b$ and $\Omega_{b}^{*}$. Phenomenological
implication of our result is discussed.

\end{minipage}
\end{center}

\vspace{1cm}

\vspace{2cm}

\baselineskip 22pt

\newpage
\noindent{\bf I. Introduction}

\vspace{0.1cm}

The present data of the lifetime ratio of $B$ and $\Lambda_b$ are
\cite{Data}
\begin{eqnarray}
\label{data}
{\tau(B^-)\over \tau(B_d)} &=& 1.06\pm 0.04, \nonumber\\
{\tau(\Lambda_b)\over\tau(B_d)} &=& 0.79\pm 0.06.
\end{eqnarray}
Deviation of the ratios from unity manifests some important issues missing
in our present theoretical framework.

Since the new experiments on heavy flavor physics have accumulated more and more
data, we have a chance to get better insight into the physics which governs
the transition processes.

The heavy quark effective theory (HQET) has achieved great success in
evaluating physical processes of heavy hadrons. One can expect that for
the processes where
heavy flavors (b and c) are involved, it is possible to establish a
reasonable framework to make more accurate calculations and predictions
\cite{Geo}. At least the leading contribution would be effectively obtained
in the suggested scenario. It is generally believed that the decay
processes at quark level occur via the heavy quark decays \cite{Ruc}. Because
the estimation of hadron lifetimes  only refers to inclusive
processes, where final states are only composed of free quarks and gluons,
all calculations are more reliable except the binding effects of the initial
hadron which may bring up some unfixed factors. In the heavy quark
limit, the lifetime is mainly
determined by the decay rate of the heavy flavors which can be accurately
calculated in the framework of weak effective Hamiltonian. If it is true,
the lifetime of $B^-$, $B_d$ and $\Lambda_b$ must be close up to some phase
space kinematics, because the leading order of the expansion with respect to
inverse powers of the b-quark mass \cite{Mano} is the same for all of
them. The first order correction $(\Lambda_{QCD}/m_b)^2$ can only result in
very small change.
However, a discrepancy between the measured data and theoretical prediction
on $\tau(\Lambda_b)/\tau(B)$ is obvious.

As a matter of fact, the observation $\tau(D^{\pm})\sim 2\tau(D^0)$ while
$\tau(B^{\pm})\sim\tau(B^0)$ attracts attentions of theorists for a long
time. Bigi et al. attributed the difference to the Pauli interference
$\Delta\Gamma_{PI}$ at quark level \cite{Bigi}. Voloshin and Shifman also
discussed similar effects for charmed mesons and baryons \cite{Vol}. They
have noticed the significance of the non-spectator effects. In their work,
$\tau(\Lambda_b)/\tau(B)$ was estimated as $\sim$0.9 which still deviates
from the present data \cite{Data}.

Usually the light flavors are treated as spectators in the transitions,
namely they do not participate in the quark level processes. Their 
existence only manifests at hadronization. 
Neubert and Sachrajda carefully studied these 
effects on the lifetime differences and other quantities, such as 
$N_c$ \cite{Neu}. Obviously, the discrepancy about the ratios implies that 
the light flavors should get involved in the weak transition. 
In most cases, the non-spectator effects would be small,
however, in some processes, they can play important roles.

As shown in most literatures, the non-spectator
effects  for B-meson exclusive decays  are small and
we have also calculated \cite{Dai} these effects in B-decays based on the
well-established weak effective Hamiltonian and found that they can only
bring up at most 3 to 4 per cent of corrections.

In this work we turn to study if the non-spectator effects can more affect
the lifetime of a b-baryon. Since a b-baryon contains a heavy b-quark and
two light quarks,
its binding structure and decay mechanism are not as certain as for
B-meson, and we may have room to conceive some possibilities which involve
the light flavors and increase the total decay width of b-baryons.

The diquark structure has been considered for a long time \cite{Ebert}.
Especially because of the extra
spin symmetry, the heavy baryons can be classified by the total spin of the
light flavors, so that a diquark structure would be quite reasonable
\cite{Mannel}. Parallel studies on the form factors of $\Lambda_b$
or $\Lambda_c$ decays have also been carried out in
some literatures \cite{Korner}.

In this work, we assume that a b-baryon is constructed by a b-quark and
a diquark made of two light quarks. The diquark may have a total spin 0 or 1
and can be seen as scalar or axial-vector boson-like particles of color
triplet. While evaluating weak transitions, an imposed postulation is needed
that the diquark can undergo a flavor transition or spin-change, but does not
dissolve during the process \cite{Anse}. In this scenario, the diquark is
somehow treated as an elementary particle.
Actually, they are not point-like, but as some authors discussed,
several form factors can be used to describe the inner structure of diquarks
phenomenologically. The reactions
involving light flavor diquarks were studied by Anselmino, Kroll and Pire
\cite{Anse}. We are going to employ the effective vertices
and concerned parameters given in \cite{Anse}
to carry out the estimations of the non-spectator
effects on b-baryon lifetimes. In the calculations, by comparing the
derived result of $\tau(\Lambda_b)$ with data, we obtain a
phenomenological parameter $\beta$, then with this value, we go on calculating
the weak decay rates for $\Sigma_{b}^{(*)}$ and $\Xi_{b}^{(*)}$ and give
predictions for lifetimes of $\Omega_b$ and $\Omega_{b}^{*}$.
Since $\Lambda_b$ and $\Omega_{b}^{(*)}$ can only decay via weak interaction,
their lifetimes are determined by weak decays. However, $\Sigma_{b}^{(*)}$ and
$\Xi_{b}^{(*)}$ can decay strongly (for instance,
$\Sigma_b\rightarrow\Lambda_b+\pi$),
their lifetimes are mainly determined by the strong decay rates.

This paper is organized as following. After the introduction, we first briefly
introduce the diquark structure of b-baryons and the
effective interaction vertices of the diquarks given in \cite{Anse}, then we
derive the formulae about the inclusive decay processes of B-meson and
b-baryons. In Sec.III, we present our numerical results for $\tau(B),
\tau(\Lambda_b)$ and make predictions on weak decay rates of $\Sigma^{(*)}_b$
and $\Xi^{(*)}_b$ as well as
$\tau(\Omega_b)$ and $\tau(\Omega^*_b)$.  Then the last
section is devoted to our conclusion and discussion.\\

\noindent{\bf II. Formulation}

\vspace{0.2cm}

(i) For $B_d$ and $B^-$ decays we can easily evaluate the non-spectator
effects in the well-established theoretical framework. The pure b-quark
decay rate for $b\rightarrow c\bar us+c\bar ud+ce\bar{\nu}_e+c\mu^-\bar{\nu}_
{\mu}+c\tau^-\bar{\nu}_{\tau}$ has been carefully evaluated by Bagan et al
\cite{Bagan} as
\begin{equation}
\Gamma_b=\Gamma(b\rightarrow c\bar us+c\bar ud)+
\sum_{l=e,\mu,\tau}\Gamma(b\rightarrow c l\bar{\nu}_l),
\end{equation}
where
\begin{eqnarray}
&& \Gamma(b\rightarrow c\bar us+c\bar ud)=(4.0\pm 0.4)\Gamma(b\rightarrow ce
\bar{\nu}_e),\nonumber \\
&& \Gamma(b\rightarrow c\tau\bar{\nu}_{\tau})=0.25\Gamma(b\rightarrow ce
\bar{\nu}_e),
\end{eqnarray}
and
\begin{eqnarray}
\Gamma(b\rightarrow ce\bar{\nu}_e) &=& |V_{cb}|^2{G_F^2m_b^5\over 192\pi^3}
[1-8({m_c\over m_b})^2 \nonumber \\
&& -12({m_c\over m_b})^4\ln{m_c^2\over m_b^2}+8({m_c\over m_b})^6-
({m_c\over m_b})^8].
\end{eqnarray}

The quark level effective Hamiltonian for these processes has been evaluated
to the next-to-leading order
QCD corrections in the SM. It is given in \cite{He} as
\begin{eqnarray}
H_{eff} = {G_F\over \sqrt{2}} [V_{f_1b}V_{f_2q}^*(c_1O^{f_1f_2}(q)
+c_2 O^{f_1f_2}(q))
-V_{tb}V_{tq}^*\sum_{i=3-6} c_i^{eff} O_i(q)],
\label{ham}
\end{eqnarray}
where
\begin{eqnarray}
O_{1}^{f_1f_2}(q)&=&\bar f_{1 \alpha}\gamma_\mu (1-\gamma_5)
f_{2 \beta} \bar q_\beta \gamma^\mu (1-\gamma_5) b_\alpha,\nonumber\\
O_2^{f_1f_2}(q) &=& \bar q \gamma_\mu (1-\gamma_5) f_2 \bar f_1 \gamma^\mu
(1-\gamma_5) b,\nonumber\\
O_{3,5}(q) &=& \bar q \gamma_\mu (1-\gamma_5) b \bar q' \gamma^\mu(1\mp \gamma_5)q',
\nonumber\\
O_{4,6}(q)&=& \bar q_{\alpha} \gamma_\mu (1-\gamma_5) b_{\beta}
\bar q'_\beta \gamma^\mu (1\mp \gamma_5) q'_\alpha.
\end{eqnarray}
In eq.(\ref{ham})
we have neglected electroweak penguin contributions which are
obviously small. $q'$ is summed over $u,d,s$ and $c$, 
$q$ can be $s$ or $d$, and
$f_i \; (i=1,2) $ can be $u$ or $c$, $\alpha$ and $\beta$ are color indices.
In our later discussions we will use the Wilson coefficients evaluated in
\cite{He}. The lifetime of $B_d$ is determined by the inclusive processes at
quark level and the binding effects are taken into account when one correctly
evaluates the transition hadronic matrix elements of $B_d\rightarrow q\bar q'$
where $q$ and $\bar q'$ are free quark and antiquark. Thus the non-spectator
processes occur via the W-boson exchange ($\bar B_d^0$) or
annihilation ($B^-$) corresponding to
$O_1$ and $O_2$ operators and the penguin-induced $O_3$ through $O_6$.
We obtain
\begin{eqnarray}
\Gamma(\bar B^0 \to c \bar u)&=&N_c|{G_F\over \sqrt{2}}V_{cb}V_{ud}^*
(c_1+{c_2\over N_c})
|^2m_c^2 f_B^2 {1\over 4} m_B(1-{m_c^2\over m_B^2})^2{1\over \pi},\nonumber\\
\Gamma(\bar B^0 \to s    \bar d)&=&N_c|{G_F\over \sqrt{2}}V_{cb}V_{cs}^*
({c^{eff}_5\over N_c} + c_6^{eff}) {2m_B^2\over m_b + m_d}|^2 f_B^2 {1\over 4}
m_B{1\over\pi},\nonumber\\
\Gamma( B^- \to s \bar c)&=&N_c|{G_F\over \sqrt{2}}V_{ub}V_{cs}^*
({c_1\over N_c} +c_2)
|^2m_c^2 f_B^2 {1\over 4} m_B(1-{m_c^2\over m_B^2})^2{1\over\pi},\nonumber\\
\Gamma( B^- \to s \bar u)&=&N_c|{G_F\over \sqrt{2}}V_{cb}V_{cs}^*
({c^{eff}_5\over N_c} + c_6^{eff}) {2m_B^2\over m_b + m_u}|^2 f_B^2 {1\over 4}
m_B{1\over\pi}.
\end{eqnarray}

Then we can write the total decay width of $\bar B^0_d$ and $B^-$ as
\begin{equation}
\Gamma(\bar B^0_d)=\Gamma_b+\Gamma_{(non)}^{\bar B^0_d};\;\;\;\;
\Gamma(\bar B^-)=\Gamma_b+\Gamma_{(non)}^{B^-},
\end{equation}
where $\Gamma_{(non)}$ denotes the non-spectator contribution to the total
width as $\Gamma_{(non)}^{\bar B^0}=\Gamma(\bar B^0\rightarrow c\bar u)
+\Gamma(\bar B^0\rightarrow s\bar d),\;
\Gamma_{(non)}^{\bar B^-}=\Gamma(\bar B^-\rightarrow c\bar d)
+\Gamma(\bar B^-\rightarrow s\bar u)$. Our numerical evaluation indicate that
such $\Gamma_{(non)}$ can only change the total width of either $\bar B^0$ or
$B^-$ by 3$\sim$ 4\% at most.\\

(ii) The effective Hamiltonian of weak interaction
for heavy quark-diquark interaction.

Since there lacks an available effective Hamiltonian for Q-D
weak interaction where
Q and D denote a heavy quark and a diquark composed of two light quarks, we
have to construct it in some reasonable way. Here we do not intend to use the
Renormalization Group Equation (RGE) to obtain a complete expression
for the  $QD\rightarrow Q'D'$ 4-body interactions, but only work at tree level.

(a) For the W-boson exchange between the heavy quark and the diquark
$$b+D\rightarrow c+D',$$
where D and D' are scalar or vector diquarks with two
light quark constituents and they reside in a color triplet. The effective
vertex at the diquarks-W boson are \cite{Anse}
\begin{eqnarray}
V_S &=& -i G_S(q_1+q_2)^{\mu}W_{\mu}, \hspace{3cm}  {\rm for}\;\; SWS' \\
V_V &=& -i[G_1(q_1+q_2)^{\mu}g^{\alpha\nu}-G_2(q_2^{\alpha}g^{\mu\nu}
+q_1^{\nu}g^{\mu\alpha}) \nonumber \\
&& +G_3(q_1+q_2)^{\mu}q_1^{\nu}q_2^{\alpha}]
\epsilon_{1\alpha}\epsilon^{*}_{2\nu}W_{\mu} \hspace{1cm} {\rm for}\;\; VWV',
\end{eqnarray}
where $S, S'$ and $V,V'$ stand for scalar and  vector diquarks, $q_1,q_2$
are momenta of $D, D'$, $\epsilon_{1\alpha}^{\lambda_1},\epsilon_{2\nu}
^{\lambda_2}$ for polarizations of axial vectors $D,D'$, respectively.
$G_S, G_1,G_2,G_3$
are form factors which were determined by fitting data. For elementary
vector particles $G_1=G_2$, $G_3=0$. According to \cite{Anse}, in our case,
the momentum transfer is small, so the $G_3$ term can be neglected and
\begin{equation}
G_S={g_s\over 2\sqrt 2}F_S(Q^2),\;\;\;\;\; G_1=G_2={g_s\over 2\sqrt 2}F_V(Q^2),
\end{equation}
where
\begin{eqnarray}
\label{form}
F_S(Q^2) &=& {\bar{\alpha}_s(Q^2)Q_0^2\over Q_0^2+Q^2}, \nonumber \\
F_V(Q^2) &=& {\bar{\alpha}_s(Q^2)Q_1^2\over Q_1^2+Q^2}, \hspace{2cm} \lambda_1=
\lambda_2=0, \nonumber\\
F'_V(Q^2) &=& {Q_2^2\over Q_2^2+Q^2}F_V(Q^2), \;\;\;\;\;\;\;\;\; {\rm otherwise}.
\end{eqnarray}
These form factors are due to the QCD interactions which bind the quarks into a
boson-like diquark. Therefore this $\bar{\alpha}_s(Q^2)$ corresponds to the effective
non-perturbative QCD coupling which is not running and takes a reasonable value
similar to that used in the potential model.

These transition amplitudes for $b(p_1)+D(q_1)\rightarrow c(p_2)+D'(q_2)$
caused by the W-boson exchange read as
\begin{eqnarray}
T_{eff}^{S} &=& {G_F\over\sqrt 2}(V_{cb}V^*_{ud})\bar c\gamma_{\mu}(1-\gamma_5)b
(q_1+q_2)^{\mu} F_S(Q^2),\;\; {\rm for\; scalar\; diquarks} \\
T_{eff}^{V} &=& {G_F\over \sqrt 2}(V_{cb}V^*_{ud})\bar c\gamma_{\mu}(1-\gamma_5)b
[(q_1+q_2)^{\mu}\epsilon_1\cdot\epsilon_2^* \nonumber \\
&& -(q_1\cdot\epsilon_2^*\epsilon_1
^{\mu}+q_2\cdot\epsilon_1\epsilon_2^{*\mu})]F_V(Q^2), \hspace{2cm} {\rm for\; vector\; diquarks}.
\end{eqnarray}

\vspace{0.2cm}

(b) The penguin-induced effective vertices.

Now we turn to the transition $b(p_1)+D(q_1)\rightarrow s(p_2)+D(q_2)$,
where a virtual gluon bridges between the quark and diquark arms, all the
formulation is similar to that for W-boson exchange, but only the coupling
$DWD'$ is replaced by $DgD$.

The vertex for $b\rightarrow s+g$ is given by Hou and Tseng \cite{Hou} as
\begin{equation}
V^a_{\mu}={G_F\over\sqrt 2}{g_s\over 4\pi^2}V_t
\bar st^a[\Delta F_1(q^2\gamma_{\mu}
-q_{\mu}\rlap /q)L-F_2i\sigma_{\mu\nu}q^{\nu}m_bR]b,
\end{equation}
with $V_t=V_{ts}^*V_{tb}$, $\Delta F_1=F_1^t-F_1^c$, $F_1^t\approx 0.25$,
$F_1^c=-{2\over 3}\ln(m_c^2/M_W^2)\approx 5.3$, $F_2\approx 0.2$.

Thus we can ignore the $F_2$ part, the transition amplitudes are
\begin{equation}
T_{eff}^{S} = {G_F\over\sqrt 2} {\alpha_s\over\pi}
V_t\bar s_it^a_{ij}\gamma_{\mu}b_jt^a_{lm}\Delta F_1
F_S(Q^2)(q_1+q_2)^{\mu},\;\; {\rm for\; scalar\; diquarks}
\\
\label{glu}
\end{equation}
\begin{eqnarray}
T_{eff}^{V} &=& {G_F\over\sqrt 2} {\alpha_s\over\pi}
V_t\bar s_it^a_{ij}\gamma_{\mu}b_j t_{lm}^a\Delta F_1
F_V(Q^2)[(q_1+q_2)^{\mu}\epsilon_{1}\cdot\epsilon_{2}^*-
(q_1\cdot\epsilon_{2}^*\epsilon_{1}
^{\mu}+q_2\cdot\epsilon_{1}\epsilon_{2}^{*\mu})],\nonumber \\
&& \hspace{7cm} {\rm for\; vector\; diquarks}.
\label{glu1}
\end{eqnarray}
Using
$$t^a_{ij}t^a_{lm}\equiv {1\over 2}[\delta_{im}\delta_{jl}-{1\over N_c}
\delta_{ij}\delta_{lm}],$$
then eqs.(\ref{glu}) and (\ref{glu1})
turn into the standard form in analog to that for
mesons.\\

(iii) The contribution of the non-spectator effects to the decay width.

$\Lambda_b$ can only decay via weak interaction to final states without bottom,
so that the weak decay rate determines its lifetime.

Because there is no convenient way to evaluate the hadronic matrix elements
like for mesons given in section (i), we can invoke a
reasonable picture. $\Lambda_b$
(or $\Sigma_b,\; \Sigma^*_b$ etc.) can "decompose" into b-quark and a diquark D
(scalar or axial vector, see next subsection), then $b(p_1)$ and $D(q_1)$ scatter
into $c(p_2)+D'(q_2)$ or $s(p_2)+D(q_2)$ while $p_1+q_1=p_2+q_2$ and
$(p_1+q_1)^2=M_{\Lambda_b}^2$. In the scattering process both initial and final
states only concern free quarks and diquarks, and
the confinement effects are reflected in
a phenomenological parameter $\beta$ in analog to the wavefunction at
origin $\psi(0)$ for the meson case. In the meson case, it is well known that
the wavefunction at origin can be related to the decay constant $f_{_B}$ as
$\psi(0)=\sqrt{M_B/6}f_{_B}$ \cite{Don}, an analog of wavefunction of $Qq$
at origin $\psi^{Qq}_{\Lambda_b}$
for baryons has been discussed by many authors \cite{Bilic}. Even
though such wavefunction at origin was employed for calculations, their values
given in literatures are quite apart from each other. Here we use the
scenario of quark-diquark, they cannot annihilate into a scalar or
vector current but a fermionic one,
and we keep $\beta$ as a free parameter to be
fixed by fitting data. The decay width of the non-spectator scattering process
can be written as
\begin{equation}
\label{wid}
\Gamma_{(non)}={1\over 2M_{\Lambda_b}}\int {d^3p_2\over (2\pi)^3}{1\over 2E_2}
{d^3q_2\over (2\pi)^3}{1\over 2\omega_2}(2\pi)^4\delta^4(p_1+q_1-p_2-q_2)
|\bar T|^2\beta^2,
\end{equation}
where $E_2$ and $\omega_2$ are the energies of the produced quark and diquark
respectively, and $\beta$ corresponds to the unknown parameter.\\

(iv) The structure of the b-baryons.

The flavor configurations of the baryons containing a
b-quark and a light color triplet subject which we identify as a diquark
have been given in \cite{Mannel},
\begin{eqnarray}
&& \Lambda_b=[(qq')_0b]_{1/2},\;\;\;\; \Xi_b'=[(qs)_0b]_{1/2},\;\;\;\;
\Sigma_b=[(qq')_1b]_{1/2},\;\;\;\; \Xi_b=[(qs)_1b]_{1/2}, \nonumber \\
&& \Omega_b=[(ss)_1b]_{1/2},\;\;\;\;\Sigma_b^*=[(qq')_1b]_{3/2},\;\;\;\;
\Xi_b^*=[(qs)_1b]_{3/2},\;\;\;\; \Omega_b^*=[(ss)_1b]_{3/2},
\end{eqnarray}
where the light quark pairs tend to constitute  scalar or
axial vector diquarks
\cite{Ebert}. If one does not distinguish the mass difference between s-quark
and $q (q')$($q (q')=u,d$), i.e., we can approximately assume the light flavor
SU(3) symmetry, then the weak decay rates of $\Sigma_{b}^{(*)}$,
$\Xi_{b}^{(*)}$ and $\Omega_{b}^{(*)}$ are the same.

By the heavy flavor $SU(2)$ symmetry, we can analyze the decay modes of b-baryons
in analog to charm-baryons. Experimental data indicate that
the mass difference of $\Omega_c$ and $\Xi_c$
is small and $M_{\Omega_c}-M_{\Xi_c}<M_K$ where $M_K$ is the kaon mass, so that
the strong decay mode $\Omega_c\rightarrow\Xi_c+K$ is forbidden by the 
kinematic constraint. From $SU(2)_f$ symmetry, one can trust that 
$M_{\Omega_b}-M_{\Xi_b}<M_K$ also
holds, therefore, such decay $\Omega_b\rightarrow\Xi_b+K$ cannot occur either.
So, in principle, $\Omega_b^{(*)}$ does not have a strong decay mode.
Moreover, since the mass splitting of $\Omega_b$ and $\Omega_b^*$ is so small
(of the order $1/m_b$), the electromagnetic transition rate between them 
should be negligible.
Therefore, the lifetimes of $\Omega_{b}^{(*)}$ are 
mainly determined by the weak interaction.
On the contrary, $\Sigma_b^{(*)}$($\Xi_b^{(*)}$) can decay into 
$\Lambda_b(\Xi^{\prime}_{b})+\pi$
via strong interaction, thus their lifetimes are determined by the strong
interaction.

As we concern the transitions $b+D\rightarrow c+D'$ or
$b+D\rightarrow s+D$, the
diquark keeps its shape after the scattering, namely we only retain the
two-body final states of quark and diquark. The case of breaking the diquark or
that the produced quark may interchange with
one of the light quarks inside the diquark
to interfere with the amplitude without breaking the diquark
is ignored. We will come to more discussions
on this issue in the last section.\\

(v) The phenomenological assumption.

The previous derivations are all based on the commonly accepted
physical picture with
well-established theories and principles, later we will make an assumption
based on our discussion given above. Namely, we attribute the lifetime
difference between B-meson and $\Lambda_b$ to the light flavor involvement,
in other words, the non-spectator scattering contributions are responsible for
the difference. Including contributions from these non-spectator
inclusive processes, we can write down the expression as
\begin{equation}
R^{th}={\Gamma_b^{(\Lambda_b)}+\Gamma_{(non)}^{(\Lambda_b)}\over
\Gamma_b^{(B)}+\Gamma_{(non)}^{(B)}}=R^{exp}={1\over 0.79}.
\end{equation}
We can obtain the phenomenological parameter $\beta$ in eq.(\ref{wid})
from this equality.

$\Lambda_b$ is composed of a b-quark and a scalar diquark, whereas
$\Sigma_{b}^{(*)}$,  $\Xi_{b}^{(*)}$, and
$\Omega_{b}^{(*)} $ are made of b and an axial-vector
diquark, and their total spins are
1/2 and 3/2 respectively. The parameter $\beta$ corresponds to the effective
wavefunction at origin in the meson case.

By SU(3) symmetry of light flavors, there can be two kinds of wavefunctions
$\psi(0)$ for the b-baryons. They correspond to spin-0 scalar and spin-1
axial vector diquarks, respectively. Furthermore, $\psi(0)\propto f$
and the coupling constant $f$ is defined as
$$<0|\eta_F|F>=f_Fu_F,$$
$$<0|\eta^{\mu}_{F^*}|F^*>=\frac{1}{\sqrt{3}}f_{F^*}u^{\mu}_{F^*},$$
where $F$, $F^*$ denote b-baryons with spin 1/2 and 3/2 respectively, $\eta_F$
and $\eta^{\mu}_{F^*}$ are proper baryonic currents, and $u^{\mu}_{F^*}$ is
the Rarita-Schwinger spinor. In the leading order of HQET, $f_F=f_{F^*}$
\cite{Groote}. With
these structures of the currents, if the light flavors are treated as a
mesonic object of color triplet, i.e., diquark, we can immediately have
$\beta\propto\psi(0)$ which is proportional to $f_{F}$ or $f_{F^*}$. The numerical
evaluations  will be done in the next section.\\

\noindent{\bf III. Numerical results}

\vspace{0.3cm}

(i) For the $\Lambda_b$ lifetime.

Since $\Lambda_b$ is composed of b and a scalar diquark $(ud)_0$, the
computation of $\Gamma_{(non)}^{(\Lambda_b)}$ is relatively easier.

Taking $m_b=4.8$ GeV, $m_c=1.5$ GeV, $M_B=5.3$ GeV, $M_{\Lambda_b}=5.6$ GeV
\cite{Data}, $V_{ts}\sim V_{cb}\approx 0.036\sim 0.042,\; V_{cs}\sim V_{tb}
\sim 1$, and $m_D=0.58$ GeV, $Q_0^2=Q_1^2=3.22$ GeV$^2$, $Q_2^2=15$ GeV$^2$
\cite{Anse} we can carry out all calculations\footnote{ In fact, there is
another parameter set
given in ref.\cite{Anse} by fitting data, as $M_D=0.7$ GeV,
$Q_0^2=Q_1^2=7.43$ GeV$^2$, $Q^2_2=20$ GeV$^2$, numerically this set produces
results which  only slightly deviate from that using the parameter set employed
in text.}.

The coupling constant at vertices
$bgb$ and $DgD$ where b and D are the b-quark and
diquark of color-triplet, is running as
\begin{equation}
\alpha_s(Q^2)={12\pi\over 23\ln (Q^2/\Lambda^2)},
\end{equation}
where $\Lambda^2$ is determined by $\alpha_s(M_Z^2)(\sim 0.12)$ \cite{Data}.
By contraries, the $\bar{\alpha}_s(Q^2)$ in the form factors of
eq.(\ref{form}) corresponds to
the non-perturbative effects of QCD, so is an effective constant as that used
in the potential model and does not need to be small. In fact, all the form factors
in eq.(\ref{form}) include this $\bar{\alpha}_s$, which can be incorporated
into the parameter $\beta$, namely $\bar{\alpha}_s$ does not show up explicitly
in all expressions, as $\beta$ is fixed by fitting data of the lifetime of
$\Lambda_b$. Thus with
$${\tau(\Lambda_b)\over\tau(B)}\sim 0.79,$$
as input, we obtain
$$\beta=0.42 \;{\rm GeV}.$$
Since $M_{\Lambda_b}<M_B+M_N$ where $M_N$ is the mass of nucleon, it has no
strong and electromagnetic decay channels. As discussed before, 
$\Omega_{b}^{(*)}$ can only decay weakly.
Thus their lifetimes are determined by
weak interaction. For $\Sigma_{b}^{(*)}$ and $\Xi_{b}^{(*)}$,
there are strong and electromagnetic
channels available, so mainly
their lifetimes are not owing to weak interactions.

According to the authors of \cite{Groote,Daiy}, $f_{\Sigma^*_b}\approx f_{
\Sigma_b}$ but
\begin{equation}
{\beta_1\over\beta}={f_{\Sigma_b}\over f_{\Lambda_b}}=1.38\sim 1.8,
\end{equation}
where $\beta_1$ is the $\beta$ parameter for $\Sigma_{b}^{(*)}$.

(ii) Weak decay rates of $\Sigma_{b}^{(*)}$ and $\Xi_{b}^{(*)}$ and lifetime
of  $\Omega_{b}^{(*)}$.

$\Sigma_b$ and $\Sigma_b^*$ are composed of a b-quark and an axial-vector 
diquark, so their wavefunctions are
\begin{eqnarray}
|\Sigma_b, 1/2, 1/2> &=& \sqrt{{1\over 3}}[\sqrt{{2\over 3}}|b,1/2,-1/2>_{i}
|D, 1,1>^i \nonumber \\
&& -\sqrt{{1\over 3}}|b,1/2,1/2>_{i}|D,1,0>^i],\\
|\Sigma_b^*, 3/2, 3/2> &=& \sqrt{{1\over 3}}|b,1/2,1/2>_{i}|D,1,1>^i,
\end{eqnarray}
where b, D refer to the b-quark and diquark of spin-1, respectively, and
"i" stands for the color index.

To calculate the cross sections of the scattering, we use the helicity-coupling
amplitude method formulated by Chung \cite{Chung}, then the rest calculations
are straightforward, even though a bit tedious. Thus we obtain
\begin{eqnarray}
{\Gamma^W(\Sigma_b)\over\Gamma(B)} &\approx  & 4.15\sim 5.02,\\
{\Gamma^W(\Sigma_b^*)\over\Gamma(B)} &\approx & 2.87\sim 3.37,
\end{eqnarray}
where $\Gamma^W$ denotes the weak decay width.

The reason of such large ratios is that the cross section for a scattering
between a spinor and a spin-1 object is one order larger than that between
a spinor and a scalar.

If this picture is right, we can expect that the inclusive weak
decay rates of $\Sigma_b$
and $\Sigma_b^*$ are $4.2\sim 5.0$ and $2.9\sim 3.4$
times larger than the B-meson total width. It is a
strong prediction. The same predictions hold for $\Xi_{b}^{(*)}$ and
$\Omega_{b}^{(*)}$ if we assume the SU(3) light flavor symmetry.
Since $\Omega_b$ and $\Omega_b^*$ decay only weakly, their
lifetimes are determined by weak decay widths. Therefore,
we predict that the lifetime ratio of $\tau (B)$ and $\tau (\Omega_b)$ is
$4.2 \sim 5.0$ and that of $\tau (B)$ and $\tau (\Omega_b^*)$ is
$2.9\sim 3.4$. The prediction range is due to the uncertainty of $\beta_1$. \\

\noindent{\bf IV. Conclusion and discussion}

\vspace{0.3cm}

(1) In this work we study the discrepancy between the measured lifetime ratio of\\
$\tau(\Lambda_b)/\tau(\bar B^0 (B^-))$ and the theoretical prediction based on
the Standard Model and assume that all the lifetime
difference is due to the non-spectator effects where the light
flavors get involved. The problem is how to correctly evaluate these
non-spectator effects.

With the assumption of a quark-diquark structure of heavy baryons which
contain a heavy quark and a diquark, the non-spectator effects for the baryons
occur via interactions between the heavy quark and the diquark. In
the case for $\bar B^0_d$ or $B^-$ mesons, the non-spectator effects are due to
interactions between the heavy quark with a light antiquark, such processes
can be evaluated in terms of the well-established weak effective Hamiltonian,
all the calculation procedures are standard and have been tested, so that the
results are relatively reliable. It is observed that a color matching factor
$(C_1+C_2/N_c)$ greatly suppresses contributions of $O_1$ and $O_2$ for $\bar
B^0_d$, while the CKM entries also wash out their contributions to $B^-$. Thus
the non-spectator effects caused by the W-boson exchange or annihilation
can be neglected in either of $\bar B^0$ and $B^-$ lifetime calculations.
Instead, we find that the penguin-induced operators $O_3\sim O_6$ can make
only 3$\sim$4\% contributions to the lifetime of both $\bar B^0_d$ and $B^-$.
In total, such non-spectator effects cannot substantially change
the lifetime of $\bar B^0_d$ and $B^-$.

The baryon structure is less understood than for mesons, indeed the
quark-diquark composition is a long-proposed picture. It is believed that
the two light quarks constitute a scalar or
axial-vector diquark of color-triplet, so it
interacts with the heavy quark via W-boson exchange or a gluon-exchange where
a penguin-effective-vertex stands for the transition of $b\rightarrow s+g$.
The interaction vertices for the quark part are perfectly described by
theories, on the contrary, the interaction forms for the diquarks are not
well established yet. Since the diquark is not a real point-like elementary
particles, even though the Lorentz structure and symmetries of the effective
interaction are completely clear in analog to that for elementary particles,
one needs to introduce some phenomenological form factors to describe the
inner structure effects of the diquark.

We take the lifetime difference of $\Lambda_b$ and $\bar B^0_d$ as input
and obtain a phenomenological parameter $\beta$ which has a physical meaning
in analog to the wavefunction at origin for meson case, namely describes
the binding effects of the b-quark and the diquark.

Taking the $\beta-$value as input which is obtained by fitting
$R=\tau(\Lambda_b)/\tau(B)\sim 0.79$,  we further make predictions on $R'=
\Gamma^W(\Sigma_b)/\Gamma(B)$ and $R"=\Gamma^W(\Sigma^*_b)/\Gamma(B)$,
and find $R'\approx 4.15\sim 5.02$, $R"\approx 2.87\sim 3.37$ respectively.
Since $\Omega_b$ and $\Omega_b^*$ decay only weakly,
these predictions show that the
lifetimes of $\Omega_b$ and $\Omega_b^*$ are about $4.2 \sim 5.0$ and
$2.9\sim 3.4$ times
shorter than that of $\bar B^0(B^-)$.

In the derivations, we use the diquark structure of baryons which is
accepted with some uncertainties. Its validity is to be confirmed or negated
by the degree of closeness of theoretical predictions to the data. That,
in fact, has been a subject of interest for quite a long time. Our work
provides a way to test the diquark structure.

Indeed, because the diquark is not a real point-like particle, phenomenological
form factors are needed. There exist several parameters in the form factors,
which in principle, should be derived from some underlying theories,
for example,
the relativistic BS equation plus perturbative and non-perturbative  QCD,
but so far none of them has been well understood, thus we would rather keep
them as parameters to be determined by fitting data. So we use the values of
the parameters $Q_0^2,Q_1^2,Q_2^2$ given in \cite{Anse} in our calculations.

Since the final states of the inclusive processes only involve free quark and
diquark, the interaction forms and the transition amplitude evaluation are
more reliable. However, we neglect possibilities that the diquark in the
initial state might be broken at the scattering process, because in that
the final state would be composed of three free quarks instead of a
quark-diquark state, the three-body final state integration would much
suppress its rate compared to two-body final state. We also neglect a
possible situation that at the process of $b+D(ud)\rightarrow s+D(ud)$ may
interfere with $b+D(ud)\rightarrow u(d)+D'(sd(u))$ where the produced s-quark
replaces a quark in the diquark. Since it refers to an interference with
breaking the diquark first, which needs to consume extra energies,
we assume that such effects are small.

All these treatments and approximations may bring up certain errors
which would cause our theoretical evaluations deviate from
real values of R' and R".
The ratios will be measured in the future. But, if the general picture
is right, one has reason to believe that the qualitative conclusion of our
results that the lifetimes of $\Omega_b,\Omega_b^*$ should be much shorter
than those of $\bar B^0,B^-$ and $\Lambda_b$. Our predictions will be
tested in the future experiments.

(2) For observational sake, we can briefly discuss the phenomenological
implication of the results.

Indeed, the branching fractions are actually an average over b-baryon weighted
by their production rates in Z decays, branching ratios and detection
coefficients. Therefore, the measured value is an admixture of $\Lambda_b,\;
\Xi_b,\;\Sigma_b,\;\Omega_b$\cite{Data}. Considering the well-accepted
mechanisms, we can assume the production rates of $\Lambda_b,\Sigma_b$
should be larger than $\Xi_b,\Omega_b$ and it is an admixture of $\Lambda_b$
and $\Sigma_b^{(*)}$. Therefore the theoretical estimation on the weak
decay rates of $\Lambda_b$ in fact is the corresponding value for the
admixture.

Our numerical results show that the weak decay rates of $\Sigma_b^{(*)}$ is
much larger than that of $\Lambda_b$, so an actual $\tau(\Lambda_b)
/\tau(B)$ may be not so serious as stated in the literatures. The problem
for $\Lambda_b$ is alleviated, but the non-spectator effects are still very
important for the b-baryons with spin-1 diquarks.

This mechanism will be tested in the future precise measurements. However,
because $\Sigma_b^{(*)}$ has strong and electromagnetic decay modes, the
complexity would make precise measurements very difficult.

First, the strong products and secondary products would contaminate the
experimental environment and  it is hard to identify if the products are
directly from the b-baryons.

Moreover, for example, $\Sigma_b$ can strongly decay into $\Lambda_b+\pi$
and then $\Lambda_b$ weakly decays into final states without bottom. It is
also hard to distinguish between the products coming directly from $\Sigma_b$
or as secondary products from $\Lambda_b$. In principle, some clever cuts
may do the job, so for example, we just measure the two-body final states
which cannot be the secondary.

Anyhow, all these measurements are rather difficult, but since the diquark
structure is very important for understanding baryons, this study becomes
interesting and necessary. Also with rapid developments of detecting techniques
and sophisticated experimental facilities, such measurements in the future
will be possible and the prospect is optimistic.
We wish, that in near future, the B-factory can
provide accurate data for the b-baryon decays and then we can testify the
diquark mechanism.\\

\vspace{2cm}

\noindent {\bf Acknowledgments:}

This work was supported in part by the National Natural Science Foundation of China
and the Australian Research Council.
One of us (Li) would like to thank Dr. X.G. He for helpful
discussions.

\end{document}